\documentclass[submission, Phys]{SciPost}

\usepackage{hyperref}
\usepackage[normalem]{ulem}
\usepackage{times,xspace}
\usepackage{graphicx,graphics,color,epsfig}
\usepackage{amsmath}
\usepackage{amssymb}
\usepackage{amsfonts}
\usepackage{amsfonts}
\usepackage{epstopdf}
\usepackage{bm}
\usepackage{hyperref}
\usepackage{longtable}     
\usepackage{url}           
\usepackage{bm}            
\usepackage{color}
\usepackage{float}
\usepackage[normalem]{ulem}

\def\be{\begin{equation}}
\def\ee{\end{equation}}
\def\bea{\begin{eqnarray}}
\def\eea{\end{eqnarray}}

\def\T{\mathcal{T}}
\def\P{\mathcal{P}}

\def\PT{\mathcal{PT}}
\def\C{\mathcal{C}}
\def\CPT{\mathcal{CPT}}
\def\Q{\mathcal{Q}}

\begin{document}


\begin{center}{\Large \textbf{
A new Berry phase term in parity-time symmetric non-Hermitian spin-1/2 quantum systems
}}\end{center}

\begin{center}
Ananya Ghatak,
Tanmoy Das$^{*}$,
\end{center}

\begin{center}
Department of Physics, Indian Institute of Science, Bangalore 560012, Karnataka, India
\\
* tnmydas@iisc.ac.in
\end{center}

\begin{center}
\today
\end{center}

\section*{Abstract}
Recently developed parity ($\mathcal{P}$) and time-reversal ($\mathcal{T}$) symmetric non-Hermitian quantum theory is envisioned to have far-reaching implications in basic science and applications. It is known that the $\PT$-inner product is defined with respect to a non-canonical, system generated dynamical symmetry, namely the $\C$ symmetry. Here, we show that the $\PT$ invariant equation of motion is defined by the simultaneous time evolution of the state $\psi(t)$ and the operator $\C(t)$ to manifest unitarity. The dynamical $\C$ operator lends itself into a new term in the Berry phase. The $\PT$ symmetric theory is not generally applicable for spin-1/2 fermions, since here $\PT$ inner product vanishes due to the Kramer's degeneracy. We consider a spin-1/2 non-Hermitian setup which acquires the combined $(\PT)^2=+1$ symmetry, despite $\T^2=-1$ and $\P^2=+1$. The Hamiltonian inherits a non-Abelian adiabatic transporter and the topological degeneracy via the combined evolution of the $\psi(t)$ state and the $\C(t)$ operator. The putative dynamical $\C$ symmetry can be a novel springboard for many other exotic quantum and topological phases.

\vspace{10pt}
\noindent\rule{\textwidth}{1pt}
\tableofcontents\thispagestyle{fancy}
\noindent\rule{\textwidth}{1pt}
\vspace{10pt}

\section{Introduction}

Topological phases in non-Hermitian (NH) Hamiltonians without and with real energy have been widely explored in recent years.\cite{GhatakJPCM} Owing to complex energy spectrum and biorthogonal probability density, one may expect a plethora of distinct topological invariants in NH systems which may or may not have any direct analog with their Hermitian counterparts.\cite{NHTITh1,NHTITh2,NHTITh3,NHTITh4,NHTITh5,NHTITh6,NHTITh7,NHTITh8,NHTITh9,NHTITh10,NHTITh11,NHTITh12,NHTITh13,NHTITh14,NHTITh15,NHTITh16,NHTITh17,NHTITh18,NHTITh19,NHTITh20,NHTITh21,NHTITh22,NHTITh23,NHTITh24,NHTITh25} Such studies can become more interesting in NH systems which posses real eigenenergies and conserved probability density with the help of pseudo-Hermitian metric,\cite{pseudo1,pseudo2,pseudo3,pseudo4,pseudo5,pseudo6,pseudo7} and/or parity ($\P$) - time reversal ($\T$) symmetry invariance.\cite{PT1,PT2,PT3,PT4,PT5} $\PT$-symmetric Hamiltonians have recently been realized in various setups (see review articles\cite{PTExpReviewNP,PTExpReviewNM}) including cold atom systems,\cite{PTOLZhang,PTOLPeng,PTFloquet} in optomechanical systems,\cite{PToptomechanics,PTTIoptomechanics} phononics,\cite{PTPhonon1,PTPhonon2} plasmonics,\cite{PTPhasmonics1,PTPhasmonics2} excitonics\cite{PTExciton},
 photonics,\cite{PTPhotonics1,PTPhotonics2,PTTIPhotonics3,PTTIQWalkLaser,PTTIQWalkPhoton1,PTTIQWalkPhoton2} metamaterials,\cite{PTMeta1,PTMeta2,PTMeta3} and others\cite{PTLCRCircuit,PTFloquet}. In fact, there have been several experimental evidence of topological phases and edge states in $\PT$ symmetric NH systems,\cite{PTTIQWalkLaser,PTTIQWalkPhoton1,PTTIQWalkPhoton2,PTTIoptomechanics,PTTIPhotonics3} and also in other generic NH systems.\cite{NHTIPhotonics,NHTIExp1,NHTIExp2}

Another motivation of the present work is to extend the $\PT$ symmetric quantum theory to {\it spinful} fermionic systems. Realization of $\PT$ invariant quantum theory in spinful NH Hamiltonians is challenging. Because the $\PT$-invariant quantum theory is generally defined for $(\PT)^2=1$, in which case the Hamiltonian and $\PT$ operator share the same eigenfunctions. For spinful systems, this is generally a problem since here $\T^2=-1$, making the $\PT$-inner product to vanish for all eigenstates (Kramer's degeneracy). 

Our work presents several novel features. 
\begin{enumerate}
\item
While $\PT$ symmetry solely can guarantee real eigenvalues, the $\PT$-inner product is not always positive and unitary.\cite{PT1,PT2,PT3,PT4,PT5}  This is remedied by the existence of a hidden, non-canonical symmetry, often termed as the $\C$ symmetry, such that the $\CPT$-inner product is positive, definite.\cite{PT4,PT5,pseudo7} Intriguingly, we show that the conservation of the probability density is defined by the {\it simultaneous} time-evolution of the state $\psi(t)$ and the $\C(t)$ operator. This situation is reminiscence of the Dirac/interaction picture, except here physical operators are not necessarily time-dependent. 

\item
We find that the adiabatic time evolution of both $\psi(t)$, $\C(t)$ conspires a new term in the Berry phase. It is also found that the gauge invariance and co-variance of the Berry phase and Berry connection, respectively, are defined with respect to a $\CPT$ invariant gauge transformation, replacing the usual unitary transformation for the Hermitian counterparts. 

\item
To incorporate {\it spinful} Hamiltonians, we consider a setup of coupled quantum wires with opposite NH spin-orbit coupling (SOC) in adjacent layers. The setup is designed to possess `parity' $\P$ via spin and sublattice inversions, such that $\P$ and $\T$ anticommutes with each other. This gives $(\PT)^2=+1$, despite $\P^2=+1$, and $\T^2=-1$. The system possesses real eigenvalues, and conserved $\CPT$ inner product under a dynamical $\C$ operator. 

\item
Under the $\CPT$ invariance, the system possess degeneracy which intrinsically governs a Berry connection, and a  winding number in a periodic boundary condition. We find that there are two distinct adiabatic transporters for the same Berry connection, which anti-commute with each other (non-Abelian), and hence gives two fold topological degeneracy. 

\item
Finally, we show that the edge state of the non-trivial topological phase has {\it harmonic oscillator} like solutions at low-energy, with complex but quantized eigenvalues and localized Gaussian wavepackets. On the other hand, owing to $\PT$ invariance in the bulk, both energy and momentum are real (conserved) in the bulk, and hence the bulk states remain delocalized - implying the absence of `skin-effect' discussed in many other NH topological systems.\cite{SkineffectAlvarez1,NHTITh22,skineffectWang1,skineffectWang2,skineffectSong,skineffectLonghi,skineffectMurakami,skineffectSato}

\end{enumerate}

\section{Dynamics of $\CPT$ - theory} 

We start with a derivation of the Berry phase in the $\CPT$-invariant quantum Hamiltonian. It is shown that the time-evolution of the eigenstates of the $\PT$-invariant Hamiltonian is governed by a different `Hamiltonian' in which the time-dependence of the $\C$-operator enters into it. The time-evolution of the $\C$-operator subsequently contributes a new term to the Berry phase, giving a novel contribution to the topological phenomena.   

\subsection{Time-dependent Schr\"odinger equation}
Let $H$ be a generic $\PT$ - invariant Hamiltonian defined as $H=(\PT)H^{T}{(\PT)}^{-1}$, where $T$ is the transpose operation. The following discussions hold for this generic requirement, however, as often done, we specialize henceforth to the symmetric ($H^{T}=H$) Hamiltonians. This implies that the specific $\PT$-invariance is defined by the commutator $[H,\PT]=0$. The Hamiltonian follows the eigenvalue equation: 
\be
H|\psi_n\rangle=E_n|\psi_n\rangle,
\label{eigen}
\ee
where eigenvalues $E_n$ are real due to $\PT$ invariance. An essential generic issue of the $\PT$-invariant quantum theory is that the $\PT$-inner product is not a constant of motion, and hence there exists a {\it dynamical}, and non-canonical symmetry $\C$ which evolves in time in such a way that the $\CPT$ - inner product becomes conserved. The $\C$  operator follows $[H,\C]=0$, and $[\PT,\C]=0$, with the $\CPT$-inner product defined as $\langle\psi^{\CPT}_m|\psi_n\rangle\equiv \langle\psi_m|\CPT|\psi_n\rangle = \int d^{d}x (\CPT\psi_m)^{T}\psi_n=\delta _{mn}$. 

For a Hermitian case, the time evolution of the state is solely governed by the Hamiltonian itself. This is not generally true for NH systems. To see that, let $\Q$ be a linear time evolution operator for the eigenstates satisfying 
\be
i\hbar |\dot{\psi}(t)\rangle = \mathcal{Q}(t) |\psi(t)\rangle,
\label{dt}
\ee
Dot denotes time-derivative. The physical constraint to keep in mind is that  the $\CPT$-inner product of the system is a constant of motion, i.e., $\partial_t \langle \psi_m | \CPT |\psi_n\rangle =0$ for all $n$ and $m$ eigenstates. This constraint dictates that
$\langle\psi_m|\dot{\C}\PT|\psi_n\rangle=-\langle\dot{\psi}_m|\CPT|\psi_n\rangle - \langle\psi_m|\CPT|\dot{\psi}_n\rangle.$
Using Eqs.~\eqref{dt} 
we obtain the equation of motion of the $\C$-operator as
\bea 
i\hbar~\dot{\C}\PT=\left[\mathcal{Q},\CPT\right].
\label{Ceqom}
\eea
(Eq.~\eqref{Ceqom} can be equivalently written as $i\hbar\dot{\C}=\left [\mathcal{Q}\C-\C\mathcal{Q}^{\PT} \right]$ where $\mathcal{Q}^{\PT}=(\PT)\mathcal{Q}(\PT)^{-1}$ is understood to be $\PT$ conjugate.) This is the equivalent of the {\it Heisenberg representation} of the time-evolution of the $\C$-operator.

Here, we can infer two properties of the $\mathcal{Q}$ operator. Firstly, Eq.~\eqref{Ceqom} implies that unlike the Hamiltonian itself, $\mathcal{Q}$ is {\it not}, in general, $\CPT$-invariant; otherwise $\C$ becomes a constant of motion. Secondly, in the above derivation we have used the symmetry property that $\mathcal{Q}^T=\mathcal{Q}$ (where $T$ is the transpose), which is in accord with the symmetric Hamiltonian assumed here. (Otherwise, the derivation can also be proceeded with keeping the transpose operator in the $\PT$ conjugation throughout the derivation below.) 

For a generic $\mathcal{Q}$ matrix, it can be expressed as a sum of a $\CPT$-symmetric matrix (say, $S$), and a $\CPT$-antisymmetric matrix (say, $A$): $\mathcal{Q}=S+A$, where $[S, \CPT]=0$, \& $\{A, \CPT\}=0$. Since either $S$ or $A$ can be chosen arbitrarily, without loosing generality, we set $S\equiv H$ for any $\CPT$-symmetric Hamiltonian. Then $A$ can be determined easily. Notice that for $A$ to be antisymmetric with $\CPT$, it should follow either (i) $[A,\PT]=0$ and $\{A,\C\}=0$, or (ii) $\{A,\PT\}=0$ and $[A,\C]=0$. In both cases, we get  from Eq.~\eqref{Ceqom} that $A=\frac{i\hbar}{2}\C^{-1}\dot{\C}$. Equating for the $\Q$ operator to be 
\be
\Q=H+\frac{i\hbar}{2}\C^{-1}\dot{\C},
\label{timeop}
\ee
 we obtain the equation of motion of the eigenstates as   
\be
i\hbar |\dot{\psi_n}\rangle = \left(H+\frac{i\hbar}{2}\C^{-1}\dot{\C}\right) |\psi_n\rangle.
\label{SE}
\ee
Eq.~\eqref{SE} reveals an interesting feature of the $\CPT$-invariant quantum systems. While $H$ is the $\PT$-invariant Hamiltonian satisfying the instantaneous eigenvalue equation Eq.~\eqref{eigen}, it does not entirely describe the time-evolution of its eigenstates. Rather, the time-dependent  Schr\"odinger equation is described by another `Hamiltonian' $\mathcal{Q}(t)$. 

We emphasize that the $\PT$-invariant Hamiltonian $H$, and the $\C$ operator need not be explicitly time-dependent. Either of them or both may have implicit time-dependence such as $\dot{\C}=\partial_{\mu} C \frac{dx^{\mu}}{dt}$, where $\mu$ stands for spatial coordinates, $x^{\mu}(t)$ is the world line (in real/momentum space as appropriate). The same for the Hamiltonian. The main point is that the Hamiltonian satisfies the instantaneous eigenvalue equation in Eq.~\eqref{eigen} within the adiabatic limit for the above theory to apply. The explicit parameter dependence of the $\C$-operator is also obtained in the $\PT$ invariant quantum theory.\cite{PT4,PTCop,pseudo7}

\subsubsection{An example}
To demonstrate $\C$ operator's roles on the time-evolution of the state, we consider a dissipative Floquet system of ultracold atoms, where the $\PT$ symmetry breaking is experimentally verified.\cite{PTFloquet} The system is a Fermi gas of $^6$Li atoms at the two lowest $^2S_{1/2}$ hyperfine levels, labeled by $|\uparrow\rangle$ and $|\downarrow\rangle$. The two spin states are coupled by a radio-frequency field with a coupling constant $J$. A resonant optical beam is coupled to one of the $|\downarrow\rangle$ state only, which introduces an imbalance between the two states at a rate of $\Gamma (t)$. The Hamiltonian for this dissipative two-spin system is given by
\be
\centering
H(t)= \left(\begin{array}{ c c }
0 &  J\\
J & -i\Gamma (t)\\
\end{array} \right).
\label{Exm1}
\ee
This Hamiltonian is not $\PT$ symmetric. However, we can decompose the Hamiltonian into a $\PT$ symmetric part $H_{\PT}=J\sigma_x + \frac{i}{2}\Gamma(t)\sigma_z$,\cite{PTFloquet} and a $\PT$ anti-symmetric part $H_{A\PT}= -\frac{i}{2}\Gamma(t)\sigma_0$ part, where $\sigma_i$ are the $2\times 2$ Pauli matrices, and $\sigma_0$ is the unit matrix. Indeed, in the experiment in Ref.~\cite{PTFloquet}, the $\PT$ symmetry breaking is observed in the $H_{\PT}$ part of the Hamiltonian, while the setup is prepared for the full Hamiltonian $H$ given in Eq.~\eqref{Exm1}. We aim to prove that the Hilbert space of $H_{\PT}$ dynamically generates the $\C$ operators which follows Eq.~\eqref{SE} such as $\frac{i\hbar}{2}\C^{-1}\dot{\C} = H_{A\PT}$.

The symmetry operators are $\P=\sigma_x$, and $\T=\mathcal{K}$ complex conjugation. The eigenvalues of $H_{\PT}$ are $E_{\pm}=\pm E$, where $E=\sqrt{J^2-\Gamma^2/4}$. The corresponding eigenvectors are
\bea 
\psi_{+} = \frac{1}{\sqrt{2EJ}}
\left(
\centering
\begin{array}{ c }
E+i\Gamma/2  \\
J\\
\end{array} 
\right),
\psi_{-} = \frac{i}{\sqrt{2EJ}}
\left(
\centering
\begin{array}{ c }
E-i\Gamma/2  \\
-J\\
\end{array} \right).
\label{Exm2}
\eea

Then $\langle \PT\psi_{\pm}|\psi_{\pm}\rangle = \pm 1$, and the $\PT$-invariant state is not normalizable. Hence we define the $\C$ operator in the usual way\cite{PT4,PT5,pseudo7} as $\C=|\PT \psi_+\rangle\langle \psi_+| + |\PT \psi_-\rangle\langle \psi_-|$ which gives $\C=\frac{1}{E}H$. So, we have $\C\psi_{\pm}=\pm \psi_{\pm}$, and $\langle \CPT\psi_{\pm}|\psi_{\pm}\rangle = 1$. Substituting for $\C$, we get $\frac{i\hbar}{2}\C^{-1}\partial_t\C=H_{A\PT}$ which satisfies Eq.~\eqref{SE}.

\subsection{Non-Abelian Berry matrix in $\CPT$ - theory} 
Let the system adiabatically evolves in time such that the Hamiltonian possess instantaneous eigenstates $|\psi_n(t)\rangle$ satisfying Eq.~\eqref{eigen} at all time. For generality, we assume each eigenstate is $N$-fold degenerate. (Since Berry phase does not dependent on onsite energy, without loosing generality, we set $E_n=0$ for simplicity. It can be shown that the result remains unchanged for $E_n\ne 0$ as long as it is real.) In the adiabatic limit, a general state $|\phi_n\rangle$ of the system can be expanded in terms of $|\psi_n(t)\rangle$ as: 
\bea
|\phi_n(t)\rangle &=&\sum_{m}|\psi_{m}(t)\rangle \Gamma_{mn}(t). 
\label{degeigenstate}
\eea
$m,n$ run over the $N$-degenerate states. Substituting $|\phi_n(t)\rangle$ in Eq.~\eqref{dt} with the time-evolution operator $\mathcal{Q}$ given in Eq.~\eqref{timeop}, we obtain
$\sum_{m}\left[|\dot{\psi}_{m}\rangle \Gamma_{mn} + |\psi_{m}\rangle \dot{\Gamma}_{mn}\right]=\frac{1}{2}\C^{-1}\dot{\C}\sum_{m}|\psi_{m}\rangle \Gamma_{mn},$
where we have substituted $H|\psi_n\rangle=0$. Multiplying $\langle \CPT\psi_{l}|$ from left, and using the orthonormal condition for the $\CPT$ inner product, we obtain 
\bea
\dot{\Gamma}_{ln} &=& -\sum_{m}\left[\left\langle\psi_{l}|\CPT|\dot{\psi}_{m}\right\rangle - \frac{1}{2}\left\langle \psi_{l}|\dot{\C}\PT|\psi_{m}\right\rangle\right] \Gamma_{mn},\nonumber\\
%
&=& i\sum_{m}\mathcal{A}_{lm} \Gamma_{mn},
\label{Udot}
\eea
where the non-Abelian Berry gauge field is defined as
\be
\mathcal{A}_{lm}=i\left[\left\langle\psi_{l}|\CPT|\dot{\psi}_{m}\right\rangle - \frac{1}{2}\left\langle\psi_{l}|\dot{\C}\PT|\psi_{m}\right\rangle\right].
\label{BerryConn}
\ee
In the case when there is no explicit time-dependence in both $H$ and $\C$, and their time-dependence is governed by the evolution in the momentum space. We obtain the Berry gauge field as
\be
\mathcal{A}_{\mu,lm}=i\left[\left\langle\psi_{l}|\CPT|\partial_{\mu}{\psi}_{m}\right\rangle - \frac{1}{2}\left\langle \psi_{l}|(\partial_{\mu}\C)\PT|\psi_{m}\right\rangle\right].
\label{BerryConnSpace}
\ee
Here we have substituted $\dot{O}=\partial_{\mu}O\frac{dx^{\mu}}{dt}$ for $O$ any operator, and, (symbolically) $\mathcal{A}dt=\mathcal{A}_{\mu}dx^{\mu}$ with $\mu$ being a space index. Notice that the second term is a new contribution arising from the adiabatic evolution of the dynamical operator $\C$. Eq.~\eqref{Udot} can be written in a matrix multiplication format as $\dot{\Gamma} = i\mathcal{A}\Gamma$ where $\Gamma$ and $\mathcal{A}$ are matrices of dimension $N\times N$. Integration of Eq.~\eqref{Udot} yields the adiabatic transporter to be 
\be
\Gamma = P e^{i\int \mathcal{A}_{\mu}dx^{\mu}}=e^{i\gamma},
\label{Umatrix}
\ee
where $P$ represents path-ordered product, in general for non-commutative Berry connections. $\gamma$ gives the non-Abelian Berry phase matrix whose components are defined by
\be
\gamma_{lm} = \int \mathcal{A}_{\mu,lm}dx^{\mu}.
\label{Berrymarix}
\ee
Clearly, $l=m$ gives the Abelian component. Correspondingly, we can define the winding number matrix as $w_{lm}=\gamma_{lm}/\pi$ in odd dimensions, and the Chern matrix as the flux of the corresponding Berry curvature in a periodic boundary condition in even dimensions.

\subsubsection{Gauge transformation and $\CPT$ co-variant derivative}
Finally, we discuss the gauge invariance condition of the $\CPT$ invariant systems. Let $|\psi'_n\rangle$ be a different choice of the eigenstate which is rotated by a $U(N)$ gauge as: $|\psi'_n\rangle = \sum_m |\psi_m\rangle U_{mn}$. The corresponding $\CPT$ conjugate is defined as $\langle \CPT\psi'_n(t)|=\sum_{m}(U^T)^{\CPT}_{mn}\langle \CPT\psi_{m}(t)|$. (Although the transpose operator on $U$ can be taken into account simply by shuffling the indices $mn$ to $nm$, however, we explicitly retain the transpose symbol for easier derivation below. In most cases $U$ is a symmetric matrix, however, we can proceed without such an assumption for generality.) Given that the inner product must be invariant  under a gauge transformation, i.e., $\langle \psi'_n|\CPT|\psi'_n\rangle=\langle \psi_n|\CPT|\psi_n\rangle$, we obtain the crucial property: 
\be
({U}^T)^{\CPT}{U}=\mathbb{I},
\label{CPTU1}
\ee
where $\mathbb{I}$ is the $N\times N$ identity matrix.\cite{foot_U} Eq.~\eqref{CPTU1} is the replacement of the unitary condition for $U$ in the Hermitian case. If $A'_{\mu}$ is the Berry connection in the new basis, then we have the gauge co-variance defined by (in matrix form):
\bea
\mathcal{A'}_{\mu}&=&({U}^T)^{\CPT}\left[\mathcal{A}_{\mu} + i\mathbb{I}\partial_{\mu}\right]{U}.
\label{BerryConnSpaceGaugeMatrix}
\eea 
All the observables ($\mathcal{O}$), such as Berry curvature, Berry rotation matrix $\Gamma$, transform under the $\CPT$ - invariant gauge transformation as $\mathcal{O}'=({U}^T)^{\CPT}\mathcal{O}{U}$. 

\begin{figure}[ht]
\centering
\includegraphics[width=0.8\columnwidth]{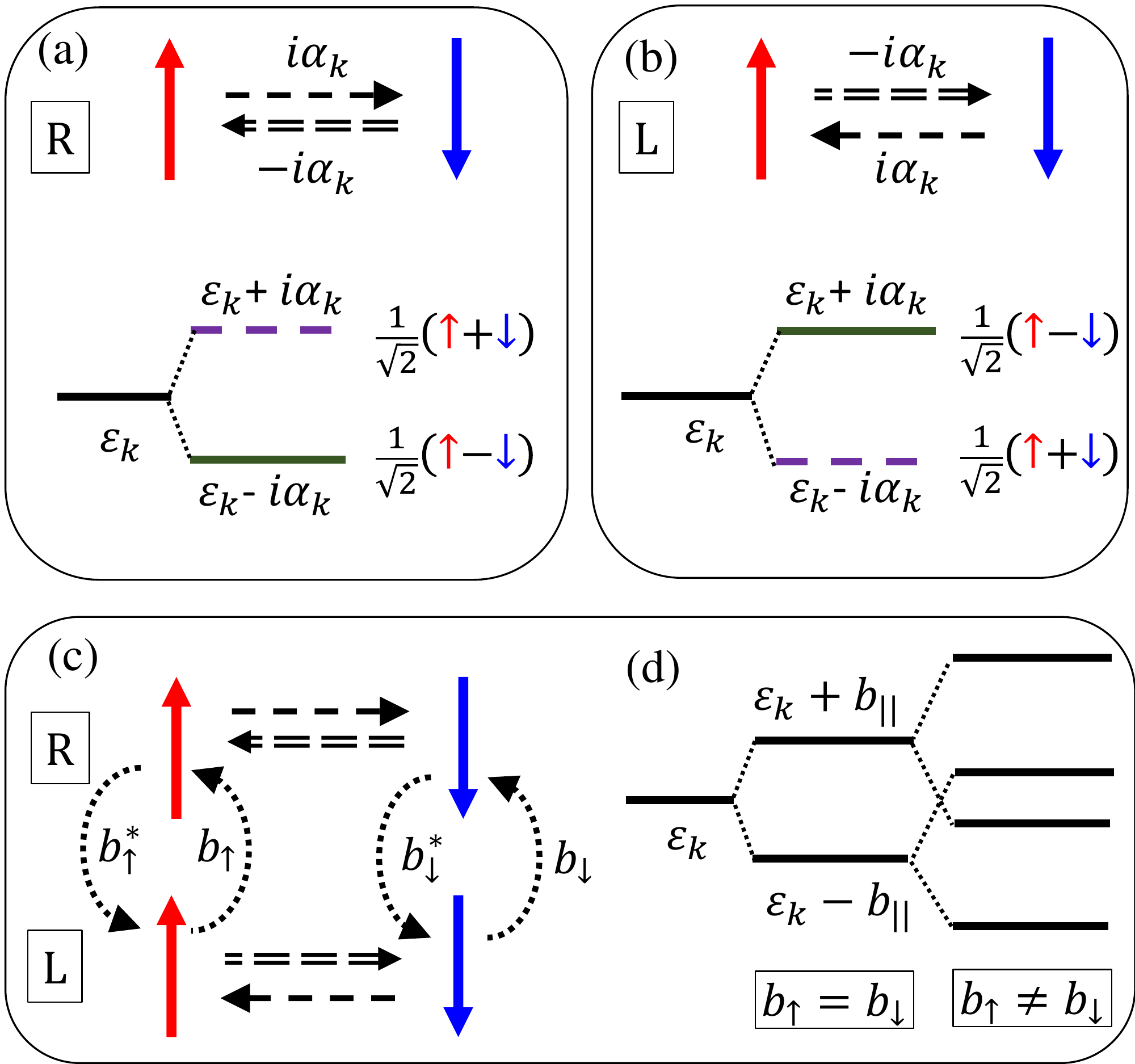}
\caption{(a) Schematic diagram of a single wire with NH SOC $i\alpha_k$, in which the hoppings between opposite spins along right and left directions are anti-Hermitian. A possible mechanism for such a NH SOC is discussed in the main text. This gives splitting of the energy levels in the imaginary plane as $\xi_{\bf k}\pm i\alpha_k$ with eigenstates as $1/\sqrt{2}(\uparrow \pm \downarrow)$, respectively. We call it right (R) wire. (b) The corresponding left (L) wire (left eigenstates of the NH R wire) is defined by the complex conjugate of the right wire in which the eigenvalues are $\xi_{\bf k}\mp i\alpha_k$ for the same eigenstates. (c) The right and left wires are coupled with spin dependent tunneling terms $b_{{\bf k},\uparrow/\downarrow}$. (d) In this setup, although the Hamiltonian is NH, but it obtains a unique $\PT$-invariance which guarantees real eigenvalues. The single band is split into two bands by $\xi_{\bf k}\pm b_{{\bf k}\parallel}$ where $b_{{\bf k}\parallel}$ (defined in the text) is real in the $\PT$ invariant region.}
\label{setup}
\end{figure}

\section{The setup} 
$\PT$-symmetric systems are generated in various setups\cite{PTExpReviewNP,PTExpReviewNM} with balanced gain and loss. The present work is directed at engineering $\PT$-symmetric spinful fermionic systems. Thus the materials search should be devoted to condensed matter systems and optical lattices of cold atoms. Recently, $\PT$-symmetric systems are successfully engineered in optically induced atomic lattices.\cite{PTOLZhang,PTOLPeng,PTFloquet}

The following setup requires a key ingredient, i.e., a NH SOC in a 1D chain of atoms. Hermitian SOC is already realized in 1D optical lattices of cold atoms,\cite{SOC1D1,SOC1D2,SOC1D3} and we propose the following feasible modification to transform it into a NH SOC. In Ref.~\cite{SOC1D1,SOC1D2,SOC1D3}, a SOC is engineered in a neutral atomic Bose-Einstein condensate with a pair of Raman lasers intersecting at $90^o$ and detuned from Raman resonance. The corresponding Hamiltonian is written as (ignoring small Zeeman coupling terms) 
\begin{equation}
h_{\bf k}=\frac{\hbar^2k^2}{2m}\sigma_{0} + \frac{\hbar^2}{m}\kappa_{\rm L}k \sigma_x,
\label{OLSOC} 
\end{equation}
where $k$, and $\kappa_{\rm L}$ are the wavevectors of the atoms and the lasers, respectively. $\sigma_x$ is the $2\times 2$ Pauli matrix defined in the spin basis of the atom, and $\sigma_{0}$ is an unit matrix. All other parameters have their usual meanings. First term gives the dispersion of the electrons while the second term gives the locking of spin with its momentum $-$ the SOC $-$ with the SOC coupling constant proportional to $\kappa_{\rm L}$.

In the setup of  Ref.~\cite{SOC1D1,SOC1D2,SOC1D3}, if the lasers are subjected to gain and/or loss following the mechanism done in, say, Refs.~\cite{PTOLZhang,PTOLPeng}, the wavevector becomes purely imaginary $\kappa_{\rm L}\rightarrow \pm i \kappa_{\rm L}$. Hence the second term $-$ the SOC term $-$ becomes anti-Hermitian. There are also other mechanism of NH SOC in the literature.\cite{NHSOC1,NHSOC2} We define SOC term as $\alpha_{\bf k} =\frac{\hbar^2}{m}\kappa_{\rm L}k$, and the dispersion $\epsilon_{\bf k}=\frac{\hbar^2k^2}{2m}$, both being real.\cite{footTB} We obtain the desired NH Hamiltonian for a single 1D atomic wire as $h_{\bf k} = \epsilon_{\bf k}\sigma_0 \pm i\alpha_{\bf k}\sigma_x$. Evidently, $\pm$ signs correspond to atomic chains with balanced gain and loss, and they are complex conjugate to each other. 

Each $h_{\rm k}$ breaks $\PT$ symmetry, with complex eigenvalues $\xi_{{\bf k},\pm}=\epsilon_{\bf k}\pm i\alpha_{\bf k}$. The corresponding right eigenvectors are $|\psi_{{\bf k},{\pm}}^{\rm R}\rangle=1/\sqrt{2}(1, \pm 1)^{T}$. The NH Hamiltonian $h_{\bf k}$ also has a corresponding left counterpart $h_{\bf k}^{\dag}$, with left eigenvectors $\langle \psi_{{\bf k},{\pm}}^{\rm L}|=(\psi_{{\bf k},{\pm}}^{\rm R})^{T}$, and left eigenvalues $\xi^*_{{\bf k},\pm}$, as shown in Fig.~\ref{setup}(b).

\subsection{The Hamiltonian of coupled $\mathcal{PT}$ - symmetric chains}
To recover the $\PT$-invariance, we propose to assemble two SOC chains with balanced gain and loss (i.e. $h_{\bf k}$ and $h_{\bf k}^{\dag}$) placed adjacent to each other such that a tunneling between them becomes active, as shown in Fig.~\ref{setup}(c). The adjacent layers are coupled by quantum tunneling amplitude $b_{k\sigma}$. The full setup is thus defined in a four-component spinor $\Psi^{T}_{\bf k}$=($\psi^{\rm R}_{{\bf k}\uparrow}$, $\psi^{\rm R}_{{\bf k}\downarrow}$, $\psi^{\rm L}_{{\bf k}\uparrow}$, $\psi^{\rm L}_{{\bf k}\downarrow}$) as 
\be
\centering
H_{\bf k}= \left(\begin{array}{ c c }
h_{{\bf k}}& V_{{\bf k}}\\
V_{{\bf k}}^{\dag} & h_{{\bf k}}^{\dag}\\
\end{array} \right).
\label{Ham}
\ee
where $V_{{\bf k}}={\rm diag}(b_{{\bf k}\uparrow},b_{{\bf k}\downarrow})$.  

The eigenvalues of Eq.~\eqref{Ham} are
\be
E_{1,3}= \epsilon \pm (b_{-} - b_{\parallel}),\quad E_{2,4}= \epsilon \pm (b_{-} + b_{\parallel}),
\label{eigenval}
\ee
where ${\bf k}$-dependence is suppressed for simplicity. $b_{{\bf k}\pm}=\frac{1}{2}(b_{{\bf k}\uparrow}\pm b_{{\bf k}\downarrow})$ and $b_{{\bf k}\parallel} = \sqrt{(b_{{\bf k}+})^2 - \alpha_{\bf k}^2}$. Since $\epsilon_{\bf k}$ and $b_{{\bf k}-}$ are always real, the $\PT$ - invariant region is simply defined by the region where $b_{{\bf k}\parallel}$ is real, i.e., $|b_{{\bf k}+}|>|\alpha_{\bf k}|$, as shown in Fig.~\ref{fig2} (lower panel). The following calculations and presentations become transparent in the complex polar coordinate defined by $b_{{\bf k}+}=b_{{\bf k}\parallel}\cosh\theta_{{\bf k}}$, and $\alpha_{\bf k}=b_{{\bf k}\parallel}\sinh\theta_{{\bf k}}$.  The corresponding eigenvectors are
\bea 
\psi_{1,3} &=& \frac{1}{\sqrt{2}}\left(
\centering
\begin{array}{ c c c c }
\sinh\frac{\theta}{2}  \\
\pm i\cosh \frac{\theta}{2}  \\
\pm\sinh \frac{\theta}{2}   \\
-i\cosh \frac{\theta}{2}  \\
\end{array} \right),
\psi_{2,4}=\frac{1}{\sqrt{2}}
\left(
\begin{array}{ c c c c }
-i\cosh{\frac{\theta}{2}} \\
\pm\sinh{\frac{\theta}{2}} \\
\mp i\cosh{\frac{\theta}{2}} \\
-\sinh{\frac{\theta}{2}} \\
\end{array} \right).
\label{eigenstates}
\eea
(${\bf k}$ dependence is implicit). Evidently, eigenvectors do not depend on $b_{{\bf k}-}$ and $\epsilon_{\bf k}$, and hence play no role on the topology. So, without loosing generality, one can set $b_{{\bf k}-}=\epsilon_{\bf k}$. The SOC is $\alpha_{\bf k}$, and the hopping term $b_{{\bf k}\sigma}$ are written in the tight-binding form as
\bea
\alpha_{\bf k}&=&\alpha_0\sin{k},\\
b_{{\bf k}\sigma}&=& b_{0\sigma} + b_{1\sigma}\cos{k},
\eea
where $\alpha_0$, and $b_{i\sigma}$ are real parameters, and $\sigma=\uparrow/\downarrow$. We choose $b_{0\sigma} = 1$, $b_{1\sigma} = -0.25$. We set $b_{1\downarrow} =0$ in Fig.~\ref{fig2}(b) to obtain the $b_{-}\ne 0$ condition. $\alpha_0=0.5$ and 1.1 give $\PT$ unbroken and broken regions, respectively. All parameter values are given in arbitrary energy unit. In Fig.~\ref{fig2} (upper panel), we plot the band dispersions for the four representative cases.

Bands $E_{1,4}$ lie below the Fermi level, while $E_{2,3}$ remain above it. For $b_{{\bf k}-}=0$, two valence bands and two conductions bands are degenerate to each other at all ${\bf k}$-points. When $b_{{\bf k}-}\ne 0$, we find that the degeneracy is lifted at all points except at two characteristic $\pm K$ points. We shall learn below that this band inversion is responsible for topological phase. Looking at the dispersions in Eq.~\eqref{eigenval}, we can easily find that $E_{1,4}$ bands do not overlap when $|b_{0\downarrow}/b_{0\uparrow}|>1$ and in such a case, the topology is lost, see Fig.~\ref{fig2}(c). In all cases, the energy spectrum exhibits particle-hole symmetry. For $\alpha_0=1.1 > b_{0+}$, the $\PT$ invariance is lost, and the energy spectrum acquires imaginary components, see Fig.~\ref{fig2}(d).

\subsection{Symmetry properties}
Its now worthwhile delineating the  symmetry properties of the Hamiltonian. In a NH, $\PT$-symmetric Hamiltonian, $(\PT)^2=+1$ is generally a requirement to ensure real eigenvalues. But for half-integer spin, we have $\T^2=-1$, and $\P^2=+1$, so that $\langle \psi_n|\PT|\psi_n\rangle=0$ for all $n$ states, and hence the $\PT$ conjugate Hilbert space collapses. 
In our Hamiltonian, both coefficients $i\alpha_{\bf k}$ and $b_{{\bf k}\sigma}$ break $\T$ symmetry; with $i\alpha_{\bf k}$ is $\PT$ invariant, while $b_{{\bf k}\sigma}$ is not (since $b_{{\bf k}\sigma}$ is even under spatial inversion). This implies that the $\P$ operator cannot be a simple chiral inversion (i.e., R$\leftrightarrow$L), but must also involve spin inversion and at the same time be unitary. Hence the total `parity' operator is a combination of momentum and spin inversions: $\P=\tau_0\otimes \sigma_x$, with $\P^2=+1$. ($\tau_i$ are Pauli matrices defined in the chiral basis with $\tau_0$ is unit matrix.) The TR operator consists of momentum inversion and spin flips: $\T=i\tau_0\otimes\sigma_y\mathcal{K}$, with $\T^2=-1$ and $\mathcal{K}$ is the complex conjugation operator. Since $\{\T,\P\}=0$, we achieve $(\PT)^2=+1$.  

For this $\PT$-operator, the $\PT$ inner product of the eigenstates of Eq.~\eqref{eigenstates} gives $\langle \psi_m|\PT|\psi_n\rangle =(-1)^{n}\delta_{mn}$. Therefore, to achieve positive, definite inner product, we define a $\C$ operator as $\C=\sum_n |\psi_n\rangle\langle \PT\psi_n|$, which gives
\be
\C_{\bf k}=(\tau_0\otimes\sigma_z)\cosh{\theta_{\bf k}} +i(\tau_y\otimes\sigma_y)\sinh{\theta_{\bf k}}.
\label{Cop}
\ee
This gives $\langle \psi_{m}|\C_{\bf k}\PT|\psi_n\rangle=\delta_{mn}$,  in which the ${\bf k}$ dependence of the $\C_{\bf k}$ operator plays a crucial role.  

Finally, we find that the Hamiltonian in Eq.~\eqref{Ham} has both the charge conjugation $\Upsilon$ and chiral $\Xi$ symmetries. We find that $\Upsilon=\tau_z\otimes\sigma_0\mathcal{K}$, and $\Xi=\Upsilon\PT$. Under these symmetries the Hamiltonian transforms as $\Upsilon H_{\bf k}\Upsilon^{-1}=-H_{-\bf k}$, and $\Xi H_{\bf k}\Xi^{-1}=-H_{\bf k}$. Both symmetries give particle-hole symmetric spectrum $\pm E_n$ ($n$ is the band index). In our case, such a condition is satisfied if the Hamiltonian is traceless, i.e. $\epsilon_{\bf k}=0$ at all ${\bf k}$-points.

\section{Topological properties}

\begin{figure}[t]
\centering
\includegraphics[width=0.99\columnwidth]{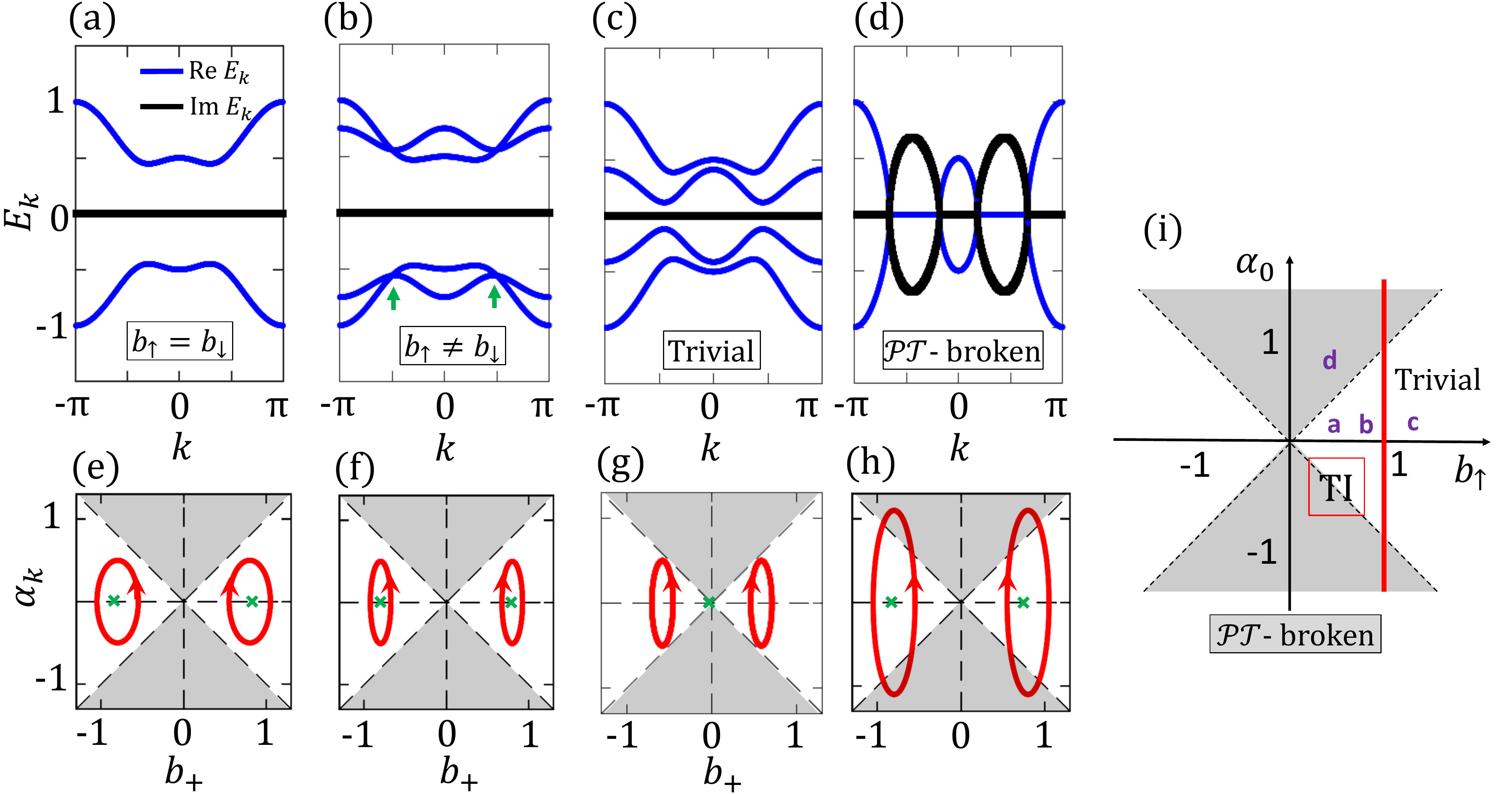}
\caption{(a-d) Band structure plots for four different representative cases: (a-c) for $\PT$-invariant and (d) for $\PT$- broken phase ($|b_{{\bf k},+}|<|\alpha_{\bf k}|$). (a),(b),(d) are in the topological phase while (c) gives a trivial phase. (a) For $b_{{\bf k}\uparrow}=b_{{\bf k}\downarrow}$, both valence and conduction band energies are real, and degenerate at all ${\bf k}$-points. (b) For $b_{{\bf k}\uparrow}\ne b_{{\bf k}\downarrow}$, both bands are split at all ${\bf k}$-points except at two characteristic ${\bf K}$ - point where the band inversion occurs (vertical green arrow). (c) A topologically trivial phase where the two valence bands do not cross at any $k$-point. (d) For $|b_{{\bf k}+}|<|\alpha_k|$, the $\PT$ symmetry is broken, and bands become complex. (e-h) We plot $b_{{\bf k}+}$ vs. $\alpha_{\bf k}$ for ${\bf k}=-\pi$ to $\pi$. This exhibit how $\theta_{\bf k}$ evolves in an adiabatic cycle for the corresponding upper panel. The green dots denote the center points $(\pm b_{K+},0)$ encircled by the contours, and the arrow on the contour dictates the sense of rotation for $k=-\pi$ to $\pi$. (i) Topological phase diagram as a function of $b_{\uparrow}$ and $\alpha_{0}$. Here  $b_{\uparrow}= b_{0\uparrow} + b_{1\uparrow}$ for $b_{\downarrow} = 1$ at ${\bf k}=0$. The gray region depicts the $\PT$ broken region, while the white region gives the $\PT$-unbroken region. The red vertical line at $0<b_{\uparrow}<1$ is the topological region, where the two valence bands cross each other. `a', `b', `c', and `d' denote the four-points where the band dispersions are drawn in (a-d), respectively.
}
\label{fig2}
\end{figure}

\subsection{Non-Abelian topology}
There exist some intriguing relationships between the odd and even eigenstates of Eq.~\eqref{eigenstates}. (i) We have $\partial_{\bf k} \psi_{1,3}=i\frac{\partial_k\theta_{\bf k}}{2}\psi_{2,4}$, and vice versa. (ii) Furthermore, $\psi_{1,3}$ and $\psi_{2,4}$ are related to each other by a rotation of $\pi/2$ and a phase of $\pi/2$. These  redundancies suggest that there exists an inherent ground state degeneracy in this system. These two conditions guarantee the existence of a robust non-Abelian Berry phase between $\psi_{1}\leftrightarrow \psi_{2}$, and $\psi_{3}\leftrightarrow \psi_{4}$ under adiabatic evolution of the state. Furthermore, with a periodic boundary condition, the corresponding winding numbers are quantized. 

The Berry connection from Eq.~\eqref{BerryConnSpace} has two terms. The adiabatic evolution of the eigenstate yields the first term:  $\mathcal{A}_{mn}^{I}=i\langle \psi_m|\C_{\bf k}\PT|\partial_{\bf k}\psi_{n}\rangle=-\frac{\partial_{\bf k}{\theta}_{\bf k}}{2}(\tau_0\otimes\sigma_x)_{mn}$. The same evolution of the $\C_{\bf k}$ operator gives $\mathcal{A}_{mn}^{II}=-\frac{i}{2}\langle \psi_m|(\partial_{\bf k}{\C}_{\bf k})\PT|\psi_{n}\rangle=\mathcal{A}_{mn}^{I}$. Hence the total Berry connection (matrix) is 
\be
\mathcal{A}=-\partial_{\bf k}{\theta}_{\bf k}(\tau_0\otimes\sigma_x).
\label{winding}
\ee
Clearly, the Berry connection is purely non-Abelian (off-diagonal), and does not have any Abelian (diagonal) component. It is interesting to note that with a rotation by $U=\tau_0 \otimes \sigma_y$, $\mathcal{A}$ can be brought to a diagonal form, however the corresponding states $U|\psi_n\rangle$ are no longer the eigenstates of the Hamiltonian. Therefore, within the adiabatic limit, $\mathcal{A}$ cannot be diagonalized. While the $\mathcal{A}$ matrices at two different ${\bf k}$ points apparently commute, but they correspond to two degenerate (adiabatic) parallel transporters or the Wilson lines with different base points. Moreover, the two adiabatic transporters anti-commute with each other, and hence give a topological degeneracy (see below).

The corresponding Berry matrix is $\gamma=(\tau_0\otimes\sigma_x)\frac{1}{2}\oint\partial_{\bf k}\theta_{\bf k}$. Clearly, the Berry phase is defined by the winding number of the phase $\theta_{\bf k}$ in the $\alpha_{\bf k}$ vs. $b_{{\bf k}+}$ plane (recall $\theta_{\bf k}=\tanh^{-1}(\alpha_{\bf k}/b_{{\bf k}+}$). For an adiabatic path from $k=-\pi$ to $\pi$ in the Brillouin zone, there are two inequivalent, non-trivial closed contours (${\bf S}^1$ space) for $\theta_{\bf k}$ in the $\alpha_{\bf k}$ vs. $b_{{\bf k}+}$ plane: one in the $b_{{\bf k}+}>0$ region and rotates clockwise, and another in the $b_{{\bf k}+}<0$ region with counter-clockwise rotation (red line in the lower panel in Fig.~\ref{fig2}). The two closed contours encircle different centers $(\pm b_{K+},0)$, respectively, where $K$ is the band inversion point [shown in Fig.~\ref{fig2}(b) by green arrow]. As long as the centers are enclosed within the contour, the contours are not simply connected (non-trivial), and belong to the homotopy group of $\pi_1({\bf S}^1)$. The winding number of each contour is associated with the $\mathbb{Z}$ group, and the Berry phase is obtained to be $\pm n\pi$ for the two contours, where $n\in \mathbb{Z}$ is the winding number. One can move the centers out of the contours by tuning, say, $b_{1\uparrow}\ne b_{1\downarrow}$ and so on, which marks a topological phase transition.

Note that the energy dispersions are insensitive to the sign of $b_{k+}$, while the eigenvectors differ by a phase of $\pi$ for $\pm|b_{k+}|$. Therefore, there is an inherent two-fold topological degeneracy in the non-trivial phase. The windings in the two contours are clockwise and counter-clockwise, respectively, which is consistent with corresponding opposite Berry phases. Since the non-trivial Berry phase is arbitrary with respect to $2\pi$, so both contours belong to the same topological class. However, important difference arises in the corresponding adiabatic transporters $\Gamma_{\pm}$ (see Eq.~\eqref{Umatrix}) for the centers $(\pm b_{K+},0)$, respectively, for the clockwise and anti-clockwise rotations. Since they encompass two different centers, and hence are not homeomorphic. The two contours are connected by a translation between the centers, and differ by a phase of $\pi$. Hence the two parallel transporters anti-commute: $\Gamma_+\Gamma_-=-\Gamma_-\Gamma_+$.  This is the origin of the non-Abelian topology in this system.

In Fig.~\ref{fig2}, we examine the topological phases at four representative cases. For $b_{{\bf k}-}=0$, in Fig. ~\ref{fig2}(a)(e), the two valence bands (and two conduction bands) are degenerate at all $k$-points. However, the band inversion occurs only at $K=\pm$ and equivalent. (Note that a band inversion occurs where two `orbital/spin' species are exchanged in a given eigenstate, which gives the Berry phase,\cite{GhatakJPCM}. This is the reason, the Berry phase is also sometimes interpreted as a statistical phase). Therefore, despite the band degeneracy at all ${\bf k}$-points, the band inversion occurs only once, and the corresponding winding number is $\pm 1$ in an adiabatic phase in the Brillouin zone. In Fig.~\ref{fig2}(b)(f), we show a case for $b_{-}\ne 0$, where the band degeneracy is lifted except at the band inversion points (marked by green arrow). In the corresponding contour in the $b_{{\bf k}+}$ vs. $\alpha_{\bf k}$ plane, the center ($b_{{\bf K}+},0)$ lies inside the contour. Hence the Berry phase remains $\pm \pi$. A topological phase transition would corresponds to a full gapping between the two valence bands, which occurs when $|b_{{\bf k}-}|>|b_{{\bf k}||}|$ in Eq.~\eqref{eigenval}. This case is shown in Fig.~\ref{fig2}(c)(g), where the band inversion points ($b_{{\bf K}+},0)$ moves outside the contour.  Finally, we illustrate a case where the $\PT$ symmetry is broken, and the contour extends to the $\PT$-broken region in the $\alpha_{\bf k}$ vs $b_{{\bf k}+}$ plane, see Fig.~\ref{fig2}(h). Here the Berry phase is finite, but complex.\cite{GhatakJPCM}


\subsection{Boundary states}
The bulk-boundary correspondence for the $\PT$-symmetric NH topological insulator is not as concrete as in the Hermitian case.\cite{GhatakJPCM,NHedge1,NHedge2,NHedge3,NHedge4} This is because even when the $\PT$-symmetry is intact in the bulk, this symmetry may be lost at the boundary for the same parameter region. In our model, the $\PT$ symmetry is broken at the boundary giving complex energy, while the bulk states remain extended. 

For the discussion of the boundary state, a suitable choice of basis is:
\bea
\phi_{{\bf k}\pm}&=&\frac{1}{2}\left(\psi_{{\bf k}\uparrow}^{\rm R} + i\psi_{{\bf k}\downarrow}^{\rm R}\right) \pm \frac{1}{2}\left( \psi_{{\bf k}\uparrow}^{\rm L} - i\psi_{{\bf k}\downarrow}^{\rm L}\right),\nonumber\\
\chi_{{\bf k}\pm}&=&\frac{1}{2}\left(i\psi_{{\bf k}\uparrow}^{\rm R} + \psi_{{\bf k}\downarrow}^{\rm R}\right)  \pm \frac{1}{2}\left(i\psi_{{\bf k}\uparrow}^{\rm L}- \psi_{{\bf k}\downarrow}^{\rm L}\right).
\eea
The rotated Hamiltonian in this basis acquires a simpler block diagonal form with two degenerate blocks for $b_{{\bf k}-}=0$ (recall that $b_{{\bf k}-}$ does not contribute to the eigenstates, and hence the topology remains intact). Hence we can seek solutions of each $2\times 2$ Block Hamiltonian which follows
\be
i\left(\begin{array}{ c c }
0 & b_{{\bf k}+} + \alpha_{\bf k}\\
-b_{{\bf k}+} + \alpha_{\bf k} & 0 \\
\end{array} \right)
\left(\begin{array}{ c}
\phi_{{\bf k}}\\
\chi_{{\bf k}}
\end{array} \right)
=E_{\bf k}
\left(\begin{array}{ c}
\phi_{{\bf k}}\\
\chi_{{\bf k}}
\end{array} \right).
\label{SE2}
\ee
(We drop the subscript $\pm$ in the states for simplicity.) We solve Eq.~\eqref{SE2} with an open boundary condition for edge state solutions in two ways. In the first part, we assume the long-wavelength limit where analytical solutions are achievable. We subsequently perform a full numerical simulation for a finite lattice to affirm the analytical results.  

\subsubsection{Continuum model} In the long-wavelength limit, the SOC term gives $\alpha_{\bf k}\rightarrow i\alpha_0\partial/\partial x$ (we set $\hbar=1$) and $b_{{\bf k}+}$ gives the domain wall potential as $b_0x$ ($\alpha_0$ and $b_0$ are real constants). By decoupling the solutions for $\phi(x)$ and $\chi(x)$ we obtain:
\bea
\left(\alpha_0^2\frac{\partial^2}{\partial x^2} + b_0^2 x^2\right)\phi(x) &=& \left(E^2+i\alpha_0b_0\right)\phi(x),\label{EOM3}\\
\left(\alpha_0^2\frac{\partial^2}{\partial x^2} + b_0^2 x^2\right)\chi(x) &=& \left(E^2-i\alpha_0b_0\right)\chi(x).\label{EOM4}
\eea
Each equation above corresponds to quantum harmonic oscillator with complex energy spectrum. The solution of Eq.~\eqref{EOM3} gives complex quantized eigenenergy and eigenstates as 
\bea
E_n &=& \omega(1+i)\sqrt{|n|},\label{Energy}\\
\phi_n(x) &=&H_n(\beta x)e^{-|\beta|^2x^2/2},\label{States}
\eea
where $n$ is {\it real} integer. (normalization is set to 1 for simplicity). $\omega=\sqrt{\alpha_0b_0}$, and $\beta=\sqrt{-\frac{b_0}{\alpha_0}}$. $H_n$ are the Hermite polynomials with real (imaginary) argument for $\alpha_0$ and $b_0$ with opposite (same) sign (i.e. when $\beta$ is real (imaginary)). In the former case, we have fully localized eigenstates $\phi_n$ at the domain walls.\cite{footnote1} Its evident that $\chi_{n}(x)=\phi_{n+1}(x)$ with eigenvalues $E_n = \omega(1+i)\sqrt{|n+1|}.\label{Energy1}$. In other words, the same $E_n$ eigenstate of both oscillators correspond to $n+1$ and $n$ eigenstates of a harmonic oscillator for $\phi$ and $\chi$ excitations.

\begin{figure}[t]
\centering
\includegraphics[width=0.8\columnwidth]{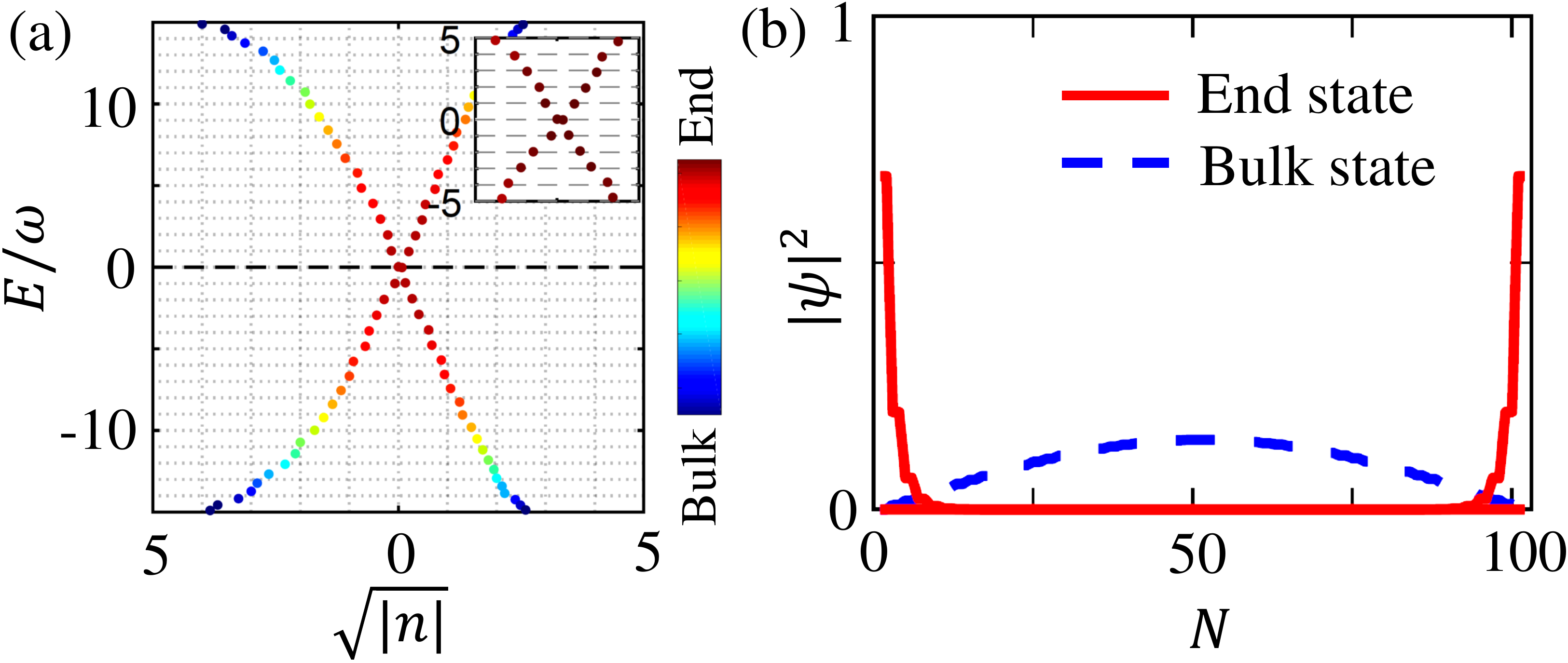}
\caption{(a) Energy ($E$) levels (normalized with $\omega$) for finite lattice model in Eq.~\eqref{LatticeModel}. The results are plotted as a function of $\sqrt{|n|}$ where $n\in \mathbb{Z}$ denotes discrete energy level, not a quantum number. We set $n=0$ for the zero energy modes. This notation is used to highlight the fact that energy levels follow harmonic oscillator like behavior at low energy [Eq.~\eqref{Energy}]. Colormap denotes the distribution of the energy levels in real space with blue to red color denote the weight from sublattice at center to the end on a particular eigenstate. For example, the zero-energy levels have red color, which means these energy levels are sitting at the end of the chain, while high-energy levels have blue color since they sit at the center of the chain. {\it Inset:} We zoom out the low energy region to highlight that the energy levels (in unit of $\omega\sqrt{|n|}$) are indeed equi-spaced with integer $n$. (b) The zero-energy states' probability density $|\psi|^2$ are plotted as a function of lattice sites. It shows that the zero modes are localized at the end of the chain (solid red line). The bulk state (higher energy states) are however extended (blue dashed line), showing the absence of the `skin-effect' in our model.
}
\label{fig:edge}
\end{figure}

\subsubsection{Lattice model with open boundary condition}
 To verify the existence of harmonic oscillator like behavior at the edge in a lattice mode, we consider a finite size lattice with $N$ atoms in each chain. Expressing Eq.~\eqref{SE2} in a finite lattice, we obtain 
\begin{eqnarray}
H &=& -\frac{1}{2}\sum_{n=1}^N\Big[b_0\Psi_n^{\dag}\sigma_y\Psi_n 
 -\Psi_n^{\dag}\left(\alpha_0\sigma_x - b_1\sigma_y\right) \Psi_{n+1}+ \Psi_{n}^{\dag}\left(\alpha_0\sigma_x + b_1\sigma_y\right) \Psi_{n-1}.
\label{LatticeModel}
\end{eqnarray}
Here $\Psi_n$ =($\phi_n$, $\chi_n$)$^T$ on the $n^{\rm th}$ site. We remind that the functional form of the inter-chain tunneling $b_{{\bf k}+}=b_0+b_1\cos{k}$, and SOC $\alpha_{\bf k}=\alpha_0\sin{k}$. We numerically diagonalize the above Hamiltonian for $N=100$ lattice sites, and the results are shown in Fig.~\ref{fig:edge}. 

In Fig.~\ref{fig:edge}(a) we notice that that the lowest energy levels are localized near the end of the 1D chains. Moreover, the low-energy energy levels are equally spaced in units of $\omega\sqrt|n|$, where $\omega$ is defined below Eq.~\eqref{Energy}, and $n$ is integer (see {\it inset} in Fig.~\ref{fig:edge}(a)). This validates the harmonic oscillator like solutions obtained in the continuum limits in Eq.~\eqref{Energy}. The higher-energy states deviate from this behavior. There are two copies of states at each end which are for the two wires.

The red-to-blue colormap on the energy levels dictates that the location of the corresponding eigenstates in real space. In Fig.~\ref{fig:edge}(b), we plot the probability density $|\psi|^2$ for the lowest energy (red solid line) and highest energy (blue dashed line) eigenstates. We find that the low-energy states are indeed localized at the two ends of the chains. However, the bulk states remain extended and follow Bloch wave nature, implying the absence of skin-effect in the present Hamiltonian.

For the case of topological `skin-effect' in other NH systems,\cite{SkineffectAlvarez1,NHTITh22,skineffectWang1,skineffectWang2,skineffectSong,skineffectSato,skineffectLonghi,skineffectMurakami} it was found that the both bulk and end states become localized at the end of the lattice. Such an effect arises from the special design of topological NH Hamiltonians where the hopping along the right and left hand directions are asymmetric.\cite{skineffectWang1,skineffectWang2} On the other hand, NH skin-effect does not occur in various other 1D models with balanced gain and loss.\cite{noSkineffectSato1,noSkineffectSato2,noSkineffectPhoton,noSkineffectLieu,noSkineffectPT,noSkineffect} Kawabata {\it et al} also demonstrated various cases where skin effect is prohibited by certain symmetries.\cite{skineffectSymm} Among the symmetries, $\PT$ symmetry of the bulk Hamiltonian plays a deciding role on the presence or absence of the skin effect and non-Bloch wavefunction. For example, a topological skin-effect is obtained for NH Hamiltonian with SOC, however, here the system does not respect $\PT$ symmetry in the bulk.\cite{skineffectSOC} In another $\PT$ symmetric NH Hamiltonian, it is shown that skin-effect arises when the $\PT$ symmetry is broken in the bulk.\cite{skineffectPTLonghi} Skin-effect is also found to be absent on a photonic $\PT$ symmetric crystal.\cite{noSkineffectPT}

The absence of skin-effect in our Hamiltonian is due to the $\PT$ invariance of the bulk Hamiltonian which guarantees that both the momentum and energy are conserved (real) for the bulk states. This is the reason, the bulk states are described by Bloch waves and remain extended. To understand it from a different point of view, we recall that although the SOC is asymmetric in a given chain, however, the SOC is reversed in the two adjacent wires. Hence in the limit of when the inter-wire tunneling is sufficiently large, the particles with finite energy can hop between the wires, and become de-localized. This is the reason, the `skin-effect' is absent here even in the topologically non-trivial phase.

\section{Conclusions}
Our work presented a number of important aspects of the $\CPT$ invariant quantum theory and model realization. 
We highlighted previously unexplored the dynamical nature of the $\C$ operator, and its crucial role on the conservation of probability in time-dependent systems, and the theory of $\CPT$ invariant gauge invariance and co-variance. The dynamical evolution of the $\C$ operator not only ensures probability conservations, but also contributes a new term to the Berry phase. We proposed a method of bypassing the hurdle of $\T^2=-1$ symmetry associated with spin-1/2 fermions to obtain the $(\PT)^2=+1$ symmetry for NH Hamiltonians with real eigenvalues. We discussed how the incorporation of balanced gain/loss in the existing technology of engineering SOC in optical lattice can generate the required NH SOC to realize our setup. We also found that the proposed spinfull Hamiltonian possess non-Abelian topological states and harmonic oscillator like localized boundary states despite complex eigenvalues. Our work contributes to the development of complete $\CPT$-invariant theory for dynamical systems, the evolution of geometric phase for spinfull systems, and sets up a pathway to achieve non-Abelian excitations without applied magnetic field or interactions.

\appendix

\section{Eigenspectrum algebra of the Hamiltonian}
The full Hamiltonian is
\be
\centering
H_{\bf k}= \left(\begin{array}{ c c c c }
\epsilon_{\bf k} & i\alpha_{\bf k} & b_{{\bf k}\uparrow} & 0\\
i\alpha_{\bf k} & \epsilon_{\bf k} & 0 & b_{{\bf k}\downarrow}\\
b^*_{{\bf k}\uparrow} & 0 & \epsilon_{\bf k} &  -i\alpha_{\bf k} \\
0  & b^*_{{\bf k}\downarrow} & -i\alpha_{\bf k} & \epsilon_{\bf k}
\end{array} \right).
\label{SHam}
\ee
The eigenvectors in terms of the dispersion spectrum are: 
\bea 
\psi_{1,3} &=& \frac{i}{\sqrt{2(d_+^2-\alpha^2)}}\left(
\centering
\begin{array}{ c c c c }
\alpha\\
\pm d_+  \\
\pm \alpha  \\
d_+\\
\end{array} \right),
\qquad\psi_{2,4} = \frac{1}{\sqrt{2(d_-^2-\alpha^2)}}\left(
\centering
\begin{array}{ c c c c }
\alpha\\
\pm d_-  \\
\pm \alpha  \\
b_-\\
\end{array} \right).
\label{eigenstates1}
\eea
(${\bf k}$ dependence is implicit above). Here we denote $b_{{\bf k}\pm}=(b_{{\bf k}\uparrow}\pm b_{{\bf k}\downarrow})/2$, and $b_{{\bf k}\parallel}=\sqrt{b_{{\bf k}+}^2-\alpha_{\bf k}^2}$, and 
$d_{{\bf k}\pm}=b_{{\bf k}+}\pm b_{{\bf k}\parallel}$. We next define a polar coordinate with complex angle as $b_{{\bf k}+}=b_{{\bf k}\parallel}\cosh\theta_{\bf k}$, and $\alpha_{\bf k}=b_{{\bf k}\parallel}\sinh\theta_{\bf k}$.  $\theta_{\bf k}=\tanh\left(\frac{\alpha_{\bf k}}{b_{{\bf k}+}}\right)$. We notice an interesting feature of the angle $\theta_k$ here. The $\PT$-symmetric region is defined by $\alpha_{\bf k}/b_{{\bf k}+}\le 1$ rending $\theta_{\bf k}\rightarrow \infty$ at the $\PT$-symmetric boundary. $\theta_{\bf k}=\pi$ at $\alpha_{\bf k}/b_{{\bf k}+}=\tanh(\pi)=0.996272$ (a transcendental number), slightly below the $\PT$ symmetric boundary. 

In this polar coordinate of $(b_{{\bf k}\parallel},\theta_{\bf k})$, we have $d_{{\bf k},+}=2b_{{\bf k}\parallel}\cosh^2\theta_{\bf k}/2$, and $d_{{\bf k},-}=2b_{{\bf k}\parallel}\sinh^2\theta_{\bf k}/2$. This gives $\sqrt{d_{{\bf k}+}^2-\alpha_{\bf k}^2}=2b_{{\bf k}\parallel}\cosh\theta_{\bf k}/2$, and $\sqrt{d_{{\bf k}-}^2-\alpha_{\bf k}^2}=2b_{{\bf k}\parallel}\sinh \theta_{\bf k}/2$. Substituting these identities in Eq.~\eqref{eigenstates1} we obtain the final expressions for the eigenstates in terms of the the angle $\theta_{\bf k}$ only as:
\bea 
\psi_{1,3} &=& \frac{1}{\sqrt{2}}\left(
\centering
\begin{array}{ c c c c }
\sinh\frac{\theta}{2}  \\
\pm i\cosh \frac{\theta}{2}  \\
\pm\sinh \frac{\theta}{2}   \\
-i\cosh \frac{\theta}{2}  \\
\end{array} \right),
\psi_{2,4}=\frac{1}{\sqrt{2}}
\left(
\begin{array}{ c c c c }
-i\cosh{\frac{\theta}{2}} \\
\pm\sinh{\frac{\theta}{2}} \\
\mp i\cosh{\frac{\theta}{2}} \\
-\sinh{\frac{\theta}{2}} \\
\end{array} \right).
\label{SMeigenstates}
\eea
For the operator  $\PT=(\tau_0\otimes \sigma_z)\mathcal{K}$ where $\mathcal{K}$ is the complex conjugation operator,  we find $(\PT\psi_n)=(\psi_n)^*$ for all all eigenstates $n$. The $\PT$ inner products are $\langle \psi_m|\PT|\psi_n\rangle=(\PT\psi_m)^T\psi_n=(-1)^n\delta_{mn}$. The $\C$ operator is defined accordingly as $\C=\sum_n |\psi_n\rangle\langle\PT\psi_n|=(\tau_0\otimes\sigma_z)\cosh{\theta_{\bf k}} +i(\tau_y\otimes\sigma_y)\sinh{\theta_{\bf k}}$. With this $\C$ operator we obtain the $\CPT$-invariant quantum theory: $\langle \psi_m| \CPT|\psi_n\rangle=\delta_{mn}$, and $\sum_n |\psi_n\rangle\langle\CPT\psi_n|=\mathbb{I}$. 

It is interesting to identify a crucial correspondence between the $\psi_{1,3}$ and $\psi_{2,4}$ eigenstates: $\partial_{\bf k}\psi_{1,3}=i\frac{\partial_{\bf k}\theta_{\bf k}}{2}\psi_{2,4}$, and  $\partial_{\bf k}\psi_{2,4}=i\frac{\partial_{\bf k}\theta_{\bf k}}{2}\psi_{1,3}$. This correspondence guarantees the existence of non-Abelian Berry gauge field $\mathcal{A}^{I}_{1,2}=\mathcal{A}^{I}_{3,4}=\frac{\partial_{\bf k}\theta_{\bf k}}{2}$. The Berry connection is a real, symmetric matrix in the $\PT$ symmetric region.

\section{Edge state calculations}
The bulk-boundary correspondence for the $\PT$-symmetric NH topological insulators is not as concrete as in the Hermitian case, and relies on systems under consideration as well as the symmetry invariance. This is quite evident because of the fact that the real eigenvalues, and conserved eigenstates in the bulk are protected by the $\PT$-symmetry (or rather the $\CPT$ symmetry), whereas such a symmetry may be inevitably lost at the boundary even if the system resides in the same parameter regime. Such a case also occurs in the present Hamiltonian. For the discussion of the boundary state, a suitable choice of basis is obtained to be as follows.
\bea
\phi_{{\bf k}\pm}&=&\frac{1}{2}\left(\psi_{{\bf k}\uparrow}^{\rm R} + i\psi_{{\bf k}\downarrow}^{\rm R}\right) \pm \frac{1}{2}\left( \psi_{{\bf k}\uparrow}^{\rm L} - i\psi_{{\bf k}\downarrow}^{\rm L}\right),\nonumber\\
%
\chi_{{\bf k}\pm}&=&\frac{1}{2}\left(i\psi_{{\bf k}\uparrow}^{\rm R} + \psi_{{\bf k}\downarrow}^{\rm R}\right)  \pm \frac{1}{2}\left(i\psi_{{\bf k}\uparrow}^{\rm L}- \psi_{{\bf k}\downarrow}^{\rm L}\right).
\eea
With a suitable choice of the spinor from the $\phi_{{\bf k}\pm}$ and $\chi_{{\bf k}\pm}$ fermions as $\Phi_{\bf k}^T=(\phi_{{\bf k}+},~\chi_{{\bf k}+},~\chi_{{\bf k}-}, \phi_{{\bf k}-})$, we construct the rotation matrix as 
\be
\centering
U= \frac{1}{2}\left(\begin{array}{ c c c c }
1 & i & 1 & -i\\
i & 1 & i & -1\\
i & 1 & -i & 1\\
1 & i & -1 & 1\\
\end{array} \right).
\label{Unitary}
\ee
This rotation to the Hamiltonian in Eq.~\eqref{SHam} yields a simpler block diagonal form 
 \bea
\centering
H' &=& \epsilon\mathbb{I} + b_-(\tau_z\otimes\sigma_0) \nonumber\\
&&+
\left(\begin{array}{ c c c c }
0 & ib_+ + i\alpha & 0 & 0\\
-ib_+ + i\alpha & 0 & 0 \\
0 & 0 & 0 & ib_+ + i\alpha\\
0 & 0 & -ib_+ + i\alpha & 0
\end{array} \right).
\label{SHam1}
\eea
(${\bf k}$ dependence in all terms above is implicit.) As in the main text, we set $\epsilon_{\bf k}=0$, and $b_{{\bf k},-}=0$ since these two terms do not contribute to the eigenstates. Hence we have a rotated Hamiltonian which is completely block diagonal defined by the last term of Eq.~\eqref{SHam1}. Again since, the block diagonal terms are exactly the same, we can seek for solution of a $2\times 2$ Schr\"odinger equation of the form:
\be
\left(\begin{array}{ c c }
0 & ib_{{\bf k}+} + i\alpha_{\bf k}\\
-ib_{{\bf k}+} + i\alpha_{\bf k} & 0 \\
\end{array} \right)
\left(\begin{array}{ c}
\phi_{{\bf k}}\\
\chi_{{\bf k}}
\end{array} \right)
=E_{\bf k}
\left(\begin{array}{ c}
\phi_{{\bf k}}\\
\chi_{{\bf k}}
\end{array} \right)
\label{SE1}
\ee
(we dropped the subscript $\pm$ in the states since the solutions are same for both). In the bulk this gives the required bands $E_{{\bf k}{\pm}}=\pm\sqrt{(b_{{\bf k}+})^2-\alpha_{\bf k}^2}$ as obtained in the main text. Here we seek for solution with the open boundary conditions. In the long-wavelength limit, the SOC term gives $\alpha_{\bf k}\rightarrow i\alpha_0\partial/\partial x$ (we set $\hbar=1$) and $b_{{\bf k}+}$ gives the domain wall potential as $b_0x$. $\alpha_0$ and $b_0$ are real constant parameters. This yields the coupled equation of motion as
\bea
\left(\alpha_0\frac{\partial}{\partial x} +ib_0x\right)\phi(x) &=& E\chi(x),\label{SEOM1}\\
\left(\alpha_0\frac{\partial}{\partial x} -ib_0x\right)\chi(x) &=& E\phi(x).\label{SEOM2}
\eea
Since $\PT$ is broken here, the energy is complex. At the exceptional point where $E=0$, we get $\phi(x) \rightarrow e^{\mp ib_0/\alpha_0 x^2}$ and $\chi(x) \rightarrow e^{\pm ib_0/\alpha_0 x^2}$, indicating that there are extended solutions at zero-energy. However, as we decouple Eqs.~\eqref{SEOM1}, and \eqref{SEOM2}, we find a series of Harmonic oscillator like solutions (shifted by a constant energy):
\bea
\left(\alpha_0^2\frac{\partial^2}{\partial x^2} + b_0^2 x^2\right)\phi(x) &=& \left(E^2+i\alpha_0b_0\right)\phi(x),\label{SEOM3}\\
\left(\alpha_0^2\frac{\partial^2}{\partial x^2} + b_0^2 x^2\right)\chi(x) &=& \left(E^2-i\alpha_0b_0\right)\chi(x).\label{SEOM4}
\eea
We substitute $\alpha^2=-1/2m$, and $b_0^2=k/2$, where $m$ and $k$ are to be thought of as mass and spring constant (with suitable dimension adjustment), respectively. We also substitute $E^2+i\alpha_0b_0=\lambda^2$. Then the connection to the Harmonic oscillator becomes vivid. Following the same quantization condition of a quantum Harmonic oscillator we have $\lambda^2=(n+\frac{1}{2})\sqrt{\frac{k}{m}}$, where $n$ is real integer. This gives the solutions of Eq.~\eqref{SEOM3} as complex quantized energy
\bea
E_n &=& \omega(1+i)\sqrt{n},\label{SEnergy}\\
\phi_n(x) &=&AH_n(\beta x)e^{-|\beta|^2x^2/2}.\label{SStates}
\eea
$\omega=\sqrt{\alpha_0b_0}$, and $\beta=\sqrt{-\frac{b_0}{\alpha_0}}$. $A$ is the normalization constant. For $\alpha_0$ and $b_0$ with opposite sign, we have perfectly localized solutions for both $\phi_n$ and $\chi_n$ states at the domain walls. 
 
Its evident that $\chi_n(x)$ states are same as $\phi_n(x)$ with corresponding eigenvalues as 
\bea
E_n &=& \omega(1+i)\sqrt{n+1}.\label{Energy1}
\eea
In other words, the $E_n$ eigenstates of both oscillators correspond to $n$ and $n+1$ states of a harmonic oscillator for $\phi$ and $\chi$ excitations.


\begin{thebibliography}{1}
\bibitem{foot_U}For the initial condition of a symmetric Hamiltonian, the $U$ operator is also symmetric, however, we carry the transpose operator explicitly for book keeping.
\bibitem{PTCop}Following the last article in Ref.~\cite{PT4,PT5}, we express the $\C$ operator as $\C=e^{-i2g}$, where $g$ is an invertible dynamical matrix. Then, for the Abelian case we have $\frac{i}{2}\C^{-1}\dot{\C} = \dot{g}$, and we obtain $\mathcal{Q}=H+\hbar\dot{g}$. Then time-dependent Sch\"odinger equation can be written as $i\hbar (\partial_t + i\partial_t g) |\dot{\psi_n}\rangle = H|\dot{\psi_n}\rangle$.
\bibitem{footTB}One can easily generalize the continuum model to a tight binding model by replacing $k\rightarrow \sin k$ and $k^2=2(1-\cos k)$.
\bibitem{footnote1} The quantum harmonic oscillator has two solutions (in addition to the Hermite polynomials): $e^{\pm \beta^2x^2}$. From real $\beta$,  $e^{-\beta^2x^2}$ is the normalizable solution, while for imaginary $\beta$, $e^{+\beta^2x^2}$ gives a normalization solution. Therefore, to accommodate both solutions, we consider $e^{-|\beta|^2x^2}$ as our general solution in Eq.~\eqref{States}.
\end{thebibliography}
\end{document}